\documentclass[aps,prd,nofootinbib,superscriptaddress,twocolumn]{revtex4}

\usepackage{graphicx}
\usepackage{amsmath}
\usepackage{amsfonts}
\usepackage{amssymb}

\newcommand{\be}{\begin{equation}}
\newcommand{\ee}{\end{equation}}
\newcommand{\br}{\begin{eqnarray}}
\newcommand{\bea}{\begin{eqnarray}}
\newcommand{\eea}{\end{eqnarray}}
\newcommand{\er}{\end{eqnarray}}
\newcommand{\ba}{\begin{array}}
\newcommand{\ea}{\end{array}}
\newcommand{\bi}{\begin{itemize}}
\newcommand{\ei}{\end{itemize}}
\newcommand{\bn}{\begin{enumerate}}
\newcommand{\en}{\end{enumerate}}
\newcommand{\bc}{\begin{center}}
\newcommand{\ec}{\end{center}}

\newcommand{\lqcd}{$\Lambda_{\mathrm{QCD}}$}

\newcommand{\beq}{\begin{equation}}
\newcommand{\eeq}{\end{equation}}

\newcommand{\abs}[1]{|#1|} 
\newcommand{\hc}[2][]{#2^{\dagger #1}} 

\newcommand{\gsim}{\lower.7ex\hbox{$\;\stackrel{\textstyle>}{\sim}\;$}}
\newcommand{\lsim}{\lower.7ex\hbox{$\;\stackrel{\textstyle<}{\sim}\;$}}

\def\mysection#1{{\bf #1.} }

\begin{document}

\title{Towards Completing the Standard Model: Vacuum Stability, EWSB and Dark Matter}

\author{Emidio Gabrielli\footnote{On leave of absence from Dipart. di Fisica, Universit\'a di Trieste, Strada Costiera 11, I-34151 Trieste, Italy}}
\affiliation{National Institute of Chemical Physics and Biophysics, R\"avala 10, 10143 Tallinn, Estonia}

\author{Matti Heikinheimo}
\affiliation{National Institute of Chemical Physics and Biophysics, R\"avala 10, 10143 Tallinn, Estonia}

\author{Kristjan Kannike}
\affiliation{National Institute of Chemical Physics and Biophysics, R\"avala 10, 10143 Tallinn, Estonia}
\affiliation{Scuola Normale Superiore and INFN, Piazza dei Cavalieri 7, 56126 Pisa, Italy}

\author{Antonio Racioppi}
\affiliation{National Institute of Chemical Physics and Biophysics, R\"avala 10, 10143 Tallinn, Estonia}

\author{Martti Raidal}
\affiliation{National Institute of Chemical Physics and Biophysics, R\"avala 10, 10143 Tallinn, Estonia}
\affiliation{Institute of Physics, University of Tartu, Estonia}

\author{Christian Spethmann}
\affiliation{National Institute of Chemical Physics and Biophysics, R\"avala 10, 10143 Tallinn, Estonia}

\date{\today}


\date{\today}

\begin{abstract}
We study the standard model (SM) in its full perturbative validity range between \lqcd\ and the $U(1)_Y$ Landau pole, assuming that a yet unknown gravitational theory
in the UV does not introduce additional particle thresholds, as suggested by the tiny cosmological constant and the absence of new stabilising physics at the EW scale.
We find that, due to dimensional transmutation, the SM Higgs potential has a global minimum at $10^{26}$ GeV, invalidating the SM as a phenomenologically
acceptable model in this energy range. We show that extending the classically scale invariant SM with one complex singlet scalar $S$ allows us to: (i) stabilise the SM Higgs potential;
(ii) induce a scale in the singlet sector via dimensional transmutation that generates the negative SM Higgs mass term via the Higgs portal;
(iii) provide a stable CP-odd singlet as the thermal relic dark matter due to CP-conservation of the scalar potential;
(iv) provide a degree of freedom that can act as an inflaton in the form of the CP-even singlet. The logarithmic behaviour of dimensional transmutation allows one to accommodate
the large hierarchy between the electroweak scale and the Landau pole, while understanding the latter requires a new non-perturbative view on the SM.

\end{abstract}

\maketitle

\section{Introduction}
The discovery~\cite{Aad:2012tfa,Chatrchyan:2012ufa} of a Higgs boson~\cite{Englert:1964et,Higgs:1964ia,Higgs:1964pj,Guralnik:1964eu} at the LHC
completes the experimental verification of the standard model (SM) as formulated in 1968 by Weinberg, Glashow and Salam \cite{Glashow:1961tr,Weinberg:1967tq, Salam:1968rm}.%
\footnote{Neutrino oscillations~\cite{Strumia:2006db} suggest that neutrinos are massive. To address this result, right handed neutrinos can be added to the SM. Whether neutrinos have Dirac or Majorana masses is yet unknown.}
This theory has passed all experimental tests during the last 40 years, leaving us without any direct evidence for new physics beyond the SM.
Indeed, all precision data, the extensive flavour physics programs at $K$- and $B$-factories and at the LHC experiments, and direct searches at LEP, the Tevatron and the LHC indicate that there are no new particles at the electroweak (EW) scale. The only possible exception is the cosmological evidence
for cold dark matter (DM) \cite{Ade:2013lta}, which likely has a particle physics origin but whose existence is known today only because of its gravitational interactions.

The fact that the only available ``new physics'' is the SM Higgs boson strongly motivates studies of its implications for our understanding of the fundamental laws of Nature.
The first set of questions to address is the phenomenological consistency of the SM itself.  Fortunately, the measured Higgs boson mass of 125 GeV turned out to
be below the SM vacuum stability bound, offering some handle to study SM inconsistencies. The second set of questions to address is what the SM inconsistencies
imply for new physics, and how to improve/extend the SM. The aim of this paper is to address both sets of questions and to formulate possible answers that can be tested in future experiments. Our attempt has a similar motivation as the previous attempt~\cite{Davoudiasl:2004be} to formulate the new SM.
We show that our present knowledge about Higgs boson properties allows us to explain EW symmetry breaking (EWSB), DM and inflation with minimal
additional degrees of freedom, with one complex scalar singlet, and with non-trivial dynamics of the model.

We would like to emphasise that in the present stage it is too early to draw any definite conclusions about the presence or absence of
new physics at the TeV scale. The LHC results from the 7-8 TeV runs gave us the Higgs and nothing else. The 14 TeV runs of the LHC may
well discover a plethora of new particles. Obviously, our interpretation of physics must follow those experimental results.
Therefore, popular scenarios of new physics like TeV scale supersymmetry or strong dynamics may well be realised in Nature.
However, we feel that it is also worthwhile and necessary to study different approaches to addressing the SM problems.
Recently it has been re-emphasised by several groups \cite{Heikinheimo:2013fta,Heikinheimo:2013xua,Farina:2013mla,Hambye:2013dgv,Giudice:2013yca,Gretsch:2013ooa,Kawamura:2013kua,Carone:2013wla,Meissner:2006zh,Meissner:2007xv,Iso:2012jn,Fujikawa:2011zf,Aoki:2012xs,Fabbrichesi:2013qca,Bazzocchi:2012de,Shaposhnikov:2007nj,Boyarsky:2009ix} that the physical naturalness argument \cite{Bardeen:1995kv}, motivated by the apparent absence of particle threshold corrections to
the cosmological constant and to the mass of the Higgs boson, allows several possibilities of formulating new physics scenarios.
In this work we start with the SM and use the principle of minimality to formulate a logically consistent view on the SM and on its potential extensions. Our approach should therefore be regarded as one logical possibility  that is subject to experimental verification in the future. In particular, we consider the possibility that no particle thresholds above the SM exist, and that gravity remains weakly coupled even for energies above the Planck scale, without significantly affecting the SM predictions. Based on these assumptions, we study the validity of the SM up to energies close to Landau pole. We will show that if this is really the case, a new vacuum instability in the Higgs potential arises, invalidating the SM theory. We will analyse a minimal extension of the SM needed to solve this problem, which consists in introducing a new complex singlet scalar field coupled to the Higgs sector. We will show that this singlet could also provide a natural DM candidate for the SM which is in well agreement with  present DM measurements.

The work plan and the main results of this paper are the following. In the next section (II) we study the running of the SM parameters in the full perturbative validity range of the SM, and show how different new physics scenarios affect our understanding of the SM properties.
In section III we assume that no high-scale particle physics thresholds exist, as suggested by present data, and show that the SM Higgs potential leads to a phenomenologically unacceptable model due to dimensional transmutation. This is a more serious problem than the metastability of the real physical vacuum, and strongly suggests that the SM must be extended in the scalar sector.
In section IV we show that by introducing a complex singlet scalar field we can understand why the universe
exists in the correct vacuum state, and how the TeV scale is generated due to dimensional transmutation.
In section V we compute the DM abundance in our model, and show that the
correct relic density can be achieved for the stable CP-odd scalar DM candidate.
In section VI we briefly discuss how cosmic inflation can be incorporated in this model,
and we conclude in section VII.

\section{The Validity  of the SM}

\begin{figure}[t]
\begin{center}
\includegraphics[width=0.45\textwidth]{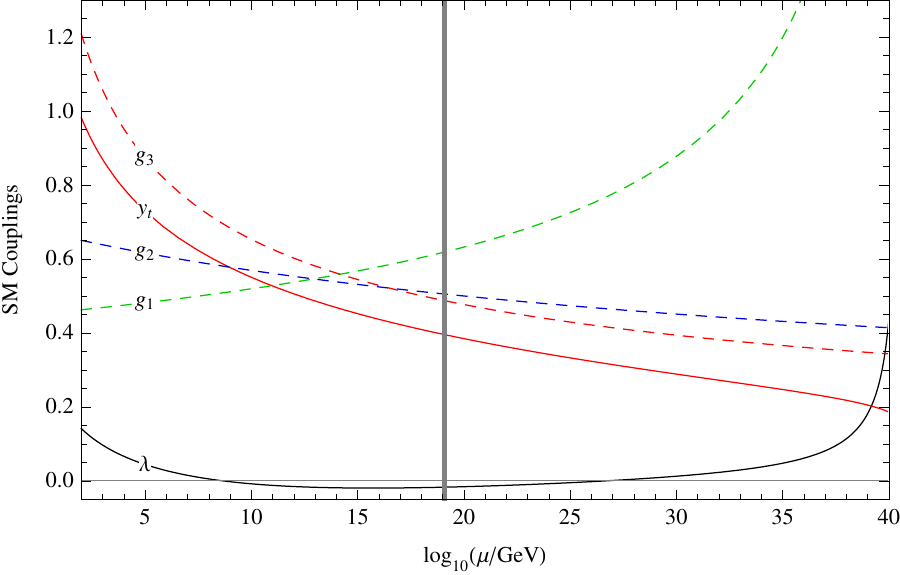}
\caption{Running of the gauge couplings, the top Yukawa and the Higgs self-coupling in the standard model. The Higgs quartic coupling is evaluated at 1-loop, the top Yukawa and the gauge couplings at 2-loop order.}
\label{fig1}
\end{center}
\end{figure}

Assuming the SM particle content and gauge symmetries, the SM as a gauge theory is technically well defined between the scale where QCD
becomes strong (\lqcd), approximately 1 GeV, and the Landau pole of the $U(1)_Y$ interaction, as depicted in Fig.~\ref{fig1}.
Below \lqcd, nature is best described by composite degrees of freedom, the mesons and nucleons. It is not known what happens at the $U(1)_Y$ Landau pole, but clearly the results of perturbation theory can no longer be trusted in the region where the $U(1)_Y$ coupling strength becomes strong. It is possible that a theory describing physics above the Landau pole would contain new degrees of freedom, and that some of the degrees of freedom of the low-energy theory are no longer useful.
Alternatively, it is possible that the degrees of freedom above the Landau pole remain the same but their dynamics must be described non-perturbatively.
In this work we accept an assumption that the existence of the Landau pole does not invalidate the SM.

The discovery of the Higgs boson fixes all the SM parameters from experimental measurements. The SM renormalisation group
equations (RGEs) are known up to 3 loops for gauge \cite{Tarasov:1980au,Larin:1993tp,Steinhauser:1998cm,Mihaila:2012fm,Mihaila:2012pz} (partially at 4-loop level for $g_3$ \cite{vanRitbergen:1997va,Czakon:2004bu}), Yukawa couplings \cite{Chetyrkin:2012rz,Bednyakov:2012en},
and the Higgs quartic coupling \cite{Chetyrkin:2013wya,Bednyakov:2013eba,Buttazzo:2013uya}. The latter computation
reduces the uncertainties related to the Higgs quartic coupling so that the biggest uncertainty in the SM RGEs is coming from the experimental
determination of the top Yukawa couplings~\cite{Degrassi:2012ry,Buttazzo:2013uya}.
In Fig.~\ref{fig1} we plot the running of top Yukawa coupling $y_t$ and the SM gauge couplings $g_1,$
$g_2,$  $g_3$ using two loop RGEs of the SM, and the Higgs boson quartic coupling $\lambda_H$ at 1-loop order. The vertical gray line shows the Planck scale.

It is astonishing that the measured SM Higgs boson mass and the other SM parameters are such that the SM Higgs potential remains perturbative
in the full validity range of the SM gauge sector. While the gauge and Yukawa couplings run significantly in this energy range of 40 orders of magnitude, the structure of the SM RGEs is such that the
Higgs quartic coupling is rather insensitive to the energy scale except close to \lqcd\  and close to the UV pole. At low energies the running of $\lambda_H$ is entirely dominated
by the running of $y_t$ which, in turn, is dominated by the running of $\alpha_s$.
At high energies the running of  $\lambda_H$ is dominated by the large value of $g_1$.
In between,  during some 25 orders of magnitude, the value of $\lambda_H$ is rather insensitive to the running of the other SM parameters, since the gauge and top Yukawa contributions have opposite
signs and cancel each other almost completely.
The measured Higgs boson mass
implies that  $\lambda_H$ runs to negative values an the intermediate scales around $10^8$~GeV, destabilising the vacuum.   The most complete studies show that
we live in the metastable vacuum very close to the critical line of vacuum decay~\cite{Degrassi:2012ry,Buttazzo:2013uya,Bezrukov:2012sa}.

The negative SM Higgs mass parameter $-\mu^2$ should be fixed from experimental data.
Its RGEs are proportional to itself, and it remains essentially constant in the full SM validity range. Due to the insensitivity to the renormalisation scale, we do not plot its
behaviour  here.

Fig.~\ref{fig1} is technically correct for the SM in isolation. Whether it is phenomenologically meaningful or not, and its potential implications, depend entirely on
which new physics completes the SM. Let us discuss the most popular scenarios going from low to high energies.

 \subsection{Strongly Interacting EWSB Scenarios}
Motivated by the analogy with chiral symmetry breaking in QCD, different strongly interacting EWSB scenarios have been proposed in the past.
Generically, all of them predict new resonances that, to explain the Higgs mass naturally, should be close to the TeV scale. Precision data and the LHC
do not support those models, although the possibility exists that they still may be realised in Nature. Due to rather restrictive experimental constraints,
it has been proposed \cite{Heikinheimo:2013fta,Hur:2011sv} that new strong dynamics generates the TeV scale in a dark sector that
is a singlet under the SM gauge group, and EWSB is then induced via a Higgs portal coupling.
If Nature has chosen the strong dynamics path, Fig.~\ref{fig1} should be terminated at $\Lambda \sim {\cal O}(1)$~TeV where the new unknown UV theory takes over.

 \subsection{Grand Unification and Supersymmetry}
Grand Unification (GUT) is perhaps the most popular new physics paradigm. It predicts a plethora of new gauge and Higgs bosons in the energy range where
the SM gauge couplings  are supposed to unify to one large gauge group, presumably around $\Lambda \sim 10^{16}$~GeV (see Fig.~\ref{fig1}). The new interactions at this scale necessarily violate baryon number, which implies proton decay through a scale-suppressed operator.
Since the GUT scale particles should induce a GUT-scale SM Higgs mass at one loop, physical naturalness requires the existence of new stabilising
particles at or below the TeV scale, such as the supersymmetric partners of the SM particles.

Although it is too early to draw definitive conclusions, negative searches for EW scale stabilising physics such as SUSY at the LHC and absence of any deviation from the
SM in flavour observables and in precision data challenge this paradigm. Should, nevertheless, GUTs exist, Fig.~\ref{fig1} would be terminated at $\Lambda \sim {\cal O}(10^{16})$~GeV where the new GUT theory should take over. In this case the SM EWSB scale must be explained by fine tuning, perhaps because of anthropic selection.

\subsection{Gravity}
Discussing how gravity affects particle physics observables is rigorously impossible for a trivial reason---no consistent and proven UV theory of gravity exists.
While there are good reasons to believe that classical general relativity should be completed by some UV theory, the Weinberg-Witten
theorem \cite{Weinberg:1980kq} strongly suggests that this theory of gravity cannot be described by a renormalisable Lorentz invariant quantum field theory (QFT) since
massless spin two particles cannot exist in those theories. Nevertheless,
the Planck scale is conventionally regarded as the scale where gravity becomes strong, depicted with a grey vertical line in  Fig.~\ref{fig1}.

There is no theoretically or experimentally supported argument that the Planck scale should be associated with the threshold of new gravitational particles.
On the contrary, the smallness of the measured cosmological constant proves that there are no tadpole contributions to the cosmological constant from any of the
known particle physics thresholds nor from hypothetical heavy new particles at the Planck scale. Therefore it is both theoretically and phenomenologically most plausible that the UV theory of gravity is very different from the known QFTs that we use to describe the SM.

Examples of this kind of theories exist. Very recently, as
a proof of concept in 2D, a new class of gravity theories was constructed \cite{Dubovsky:2013ira} which cannot be described with a local Lagrangian in the UV.
In those theories no new  particles can be associated with the scale where the classical theory is superseded, in agreement with the absence of their
contributions to the cosmological constant. Alternatively, the idea of asymptotic safety can be employed to construct theories of gravity that interacts sufficiently weakly with particle physics
\cite{ASweinberg, ASreviews1,ASreviews2,ASreviews3,ASreviews4}.

Here we do not want to prefer one approach to the theory of gravity to another. Among the two logical possibilities, that the UV theory of gravity possesses a high threshold interacting strongly with the SM or that the UV theory of gravity does not possess thresholds nor interact with the SM strongly, we adopt the second option and assume that gravity does not significantly affect the SM predictions above the Planck scale. If this assumption turns out to be wrong, all of our  model building results in this work remain correct, but the running in Fig.~\ref{fig1} must be terminated at $M_{\rm Planck}$ affecting our motivation.

\section{Phenomenological Inconsistencies of the SM Alone}

We assume here, as a theoretical possibility, that there are no new heavy particle thresholds above the EW scale. Moreover, we require  that  the SM alone is valid up to the high energy scale where the Landau pole associated to the $U(1)_Y$ gauge coupling appears. Under these hypotheses we are going to check if the SM Higgs potential generates a global minimum above the EW scale. We will  see that this will be case for energies above the Planck scale, but below the $U(1)_Y$ Landau pole. Then, we will explore the possibility to complete the SM in order to remove this unwanted minimum by the most simple generalization of the model.

\begin{figure}[t]
\begin{center}
\includegraphics[width=0.45\textwidth]{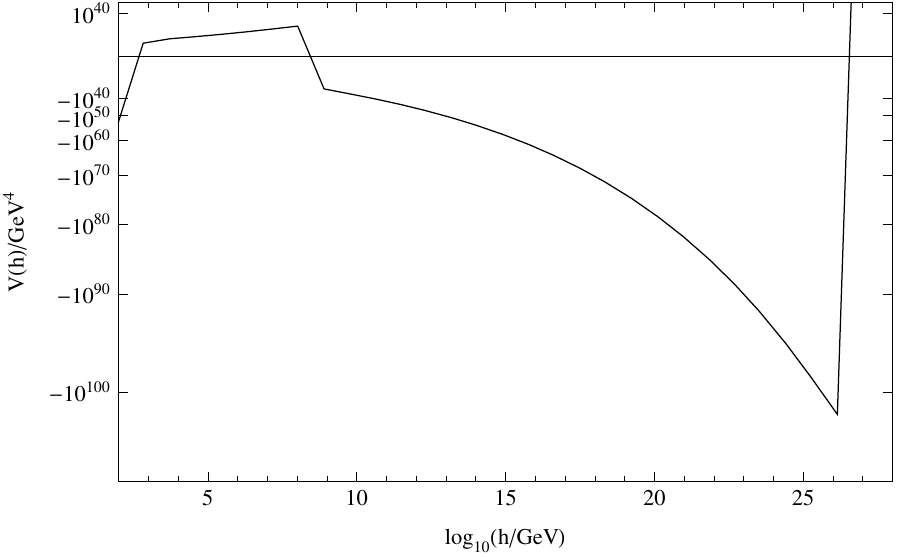}
\caption{The SM Higgs effective potential as a function of the Higgs field strength, $V(h)=-\mu^2 h^2 + \lambda_H(h) \, h^4$. The Higgs mass parameter is approximated as a constant and the running quartic coupling is evaluated at 1-loop level, where the scale is set by the field strength $h$. The global minimum at $\sim 10^{26}$ GeV is generated by $\lambda_H$ running positive (from low to high energy) at this scale. The local minimum at the electroweak scale is caused by the negative Higgs mass parameter.
}
\label{fig2}
\end{center}
\end{figure}

The first obvious task is to study the full SM Higgs effective potential, plotted in Fig.~\ref{fig2}. There is a global minimum of the potential
in the vicinity of the scale $\sim 10^{26}$~GeV where the Higgs quartic coupling $\lambda_H$ runs negative. This is a typical example of dimensional transmutation and is exactly what is expected for $\lambda_H$ to occur if its RGEs are dominated by bosonic degrees of freedom (the Higgs itself and
the gauge bosons) that run  $\lambda_H$ only toward negative values (from high to low scale).  However, the SM contains also the top quark with its large Yukawa coupling to the Higgs boson.
As explained in the last section, at low energies near the EW scale the top Yukawa becomes large and starts dominating the $\lambda_H$ running, pushing it back to positive values.
In the SM the second local minimum at low energies is obtained by adding an explicit negative Higgs mass term $-\mu^2$ to the Lagrangian.
All experiments show that we live in the low energy local minimum.

This behaviour raises three questions. The first is: What is the lifetime of our metastable vacuum? The answer to this question is already given~\cite{Degrassi:2012ry}.
The lifetime exceeds the age of the Universe and, therefore, does not
disprove the SM as a valid phenomenological theory. However,  we live dangerously close to the critical line of vacuum decay.

The second question is: What is the mechanism choosing the SM to live in the low energy local minimum instead of the global one? This is a much more serious question than
the previous one. According to our scenario the SM is understood as a low energy theory that is valid
below the $U(1)_Y$ Landau pole at $10^{40}$~GeV. The effective potential of the Higgs is generated by dimensional transmutation below the scale where the perturbative SM is
definitely valid.
Thus the SM vacuum should choose to live in the global minimum, and the vacuum expectation value (VEV) of the Higgs boson
should naturally be $\sim 10^{26}$~GeV. This is not phenomenologically acceptable.
Therefore the SM alone is not a phenomenologically acceptable theory.
This is a non-trivial result obtained only if the full SM validity range is considered. If one terminates studies of the SM Higgs potential at the GUT or Planck scale,
this result cannot be obtained.

The third question is what is the origin of the explicit Higgs mass term $-\mu^2$ and why it is so much smaller than any natural scale of the theory?
 This question can be generalised to asking whether dimensionful operators are allowed at all in the fundamental Lagrangians of physical QFTs.
 This question is also addressing the origin of the Higgs mass hierarchy problem.
  It is suggested in \cite{Heikinheimo:2013fta} that the most economical way to answer those questions is to impose classical scale invariance as a fundamental symmetry of the Lagrangian.
This automatically guarantees  the renormalizability of the theory since all higher order operators are forbidden in the Lagrangian.
Consequently, all irrelevant operators must be generated by some physical scale in the theory. This also forbids all relevant operators in the Lagrangian.
In the SM there is just one such operator, the Higgs mass term $\mu^2 |H|^2.$ Consequently, all relevant operators and all mass scales in the theory
must be generated via dimensional transmutations. Since the latter depend logarithmically on the energy scale, the existence of large hierarchies can be addressed in QFTs.
Together those ingredients can be used to explain the puzzling features of the SM. For previous work on generating the EW scale via dimensional transmutation in classically scale invariant extensions of the SM see \cite{Hambye:2013dgv,Carone:2013wla,Hempfling:1996ht,Chang:2007ki,Foot:2007ay,Foot:2007as,Foot:2007iy,Foot:2010av,Englert:2013gz,Khoze:2013uia,Dermisek:2013pta,Khoze:2013oga,Chun:2013soa,Barenboim:2013wra}, and references therein.

\section{Completing the SM with a Complex Scalar Singlet}

We showed in the last section that the SM scalar sector, as it stands today, is phenomenologically unacceptable.
The important point is that this result tells us that the SM must be improved in the Higgs sector and it also tells us how to improve it.
The physically unacceptable global minimum in the effective potential of the Higgs in Fig.~\ref{fig2} must be removed together with the explicit Higgs mass term
at low energy. The EWSB breaking scale must be obtained via dimensional transmutation from the UV Landau pole, which allows us to address the
hierarchies of the SM. If this procedure induces also the correct amount of DM our goals are achieved and the SM is completed.

\subsection{SM-like EWSB via Dimensional Transmutation}

Those tasks can all be achieved by extending the SM particle content with one complex scalar singlet field $S.$
We assume that the theory is classically scale invariant, allowing us to generate the phenomenologically
observed scales with dimensional transmutation. In our framework this implies that the scalars of the improved
SM, the Higgs doublet $H$ and the singlet $S$, must be exactly massless at tree level. As we will show, the DM is stable due to CP conservation of the scalar potential.

As we saw in the previous section, the SM Higgs develops a VEV of the order of $\sim 10^{26}$ GeV via dimensional transmutation. This happens because the Higgs self coupling $\lambda_H$ becomes negative at that scale, when running from the UV towards IR. This destabilises the tree level potential and therefore generates a minimum in the effective potential around the scale where the coupling crosses zero. However, if $\lambda_H$ were to cross zero around the TeV scale instead of the high scale at $10^{26}$ GeV, the EW symmetry breaking VEV could be generated in this manner. Since this can not be achieved with the SM couplings, the simplest solution is to add a singlet scalar $S$, and fix the couplings of $S$ so that its self coupling $\lambda_S$ crosses zero at a suitable scale, generating a VEV for $S$. This VEV can then be mediated to the SM Higgs via the portal coupling
\begin{equation}
\lambda_{SH}|S|^2|H|^2.
\label{eq:portalcoupling}
\end{equation}
If the sign of the portal coupling is negative, the Higgs gets a negative mass term from the VEV of $S$ and breaks the electroweak symmetry as in the SM.

The portal term (\ref{eq:portalcoupling}) also affects the value of the Higgs boson quartic coupling $\lambda_H$. There are two known effects.
First, the running of  $\lambda_H$ is modified by additional bosonic contributions to the RGEs so  that it may never cross  zero~\cite{Kadastik:2011aa},
and may stay positive in the whole range between \lqcd\ and the $U(1)_Y$ Landau pole $\Lambda_{\rm UV}$. Then the global minimum around $10^{26}$ GeV would not exist and the EW symmetry breaking vacuum could be naturally understood as the dynamically generated global minimum of the effective potential.
The second effect is a positive contribution to $\lambda_H$ by integrating out a scalar with a VEV~\cite{EliasMiro:2012ay,Lebedev:2012zw}.
We show in this work that the latter is numerically negligible and only the first
mechanism can be used to save the vacuum.

However, for the portal term to have a large enough effect on the running of $\lambda_H$ to keep it positive in the whole perturbative range of the SM, the portal coupling $\lambda_{SH}$ has to be large, and this will induce a large mixing between the singlet $S$ and the Higgs, implying large deviations from the SM values for the Higgs couplings. Thus this scenario is heavily constrained by the LHC data.%
\footnote{For a scenario where a large mixing could be experimentally allowed see \cite{Heikinheimo:2013cua}.}

As we will show below, this problem can be solved by making the singlet $S$ a complex field
with explicitly broken global $U(1)$ symmetry. Then there are two new degrees of freedom, the real and imaginary parts of the field, $s_R$ and $s_I$, and several couplings between these fields and the Higgs that can be used to remove the minimum at $10^{26}$ GeV while keeping the mixing effects small. Additionally, due to a residual $\mathbb{Z}_2$ symmetry, the imaginary part $s_I$ will be stable and can be interpreted as the DM particle.
For another study of a complex singlet scalar with a different motivation see \cite{Barger:2008jx}.

\subsection{The Effective Scalar Potential}

The most general scalar potential invariant under the SM gauge group and the CP transformation\footnote{The potential (\ref{eq:V:SM:singlet}) is in fact symmetric separately under the CP transformation of the SM Higgs, $H \to H^\dagger$, and the $\mathbb{Z}_2$ transformation of the singlet $S \to S^\dagger$. For notational convenience we label these both as CP, and call the real part of the singlet $s_R$, which is even under the $\mathbb{Z}_2$ transformation, {\it CP-even}, and the imaginary part $s_I$, which is odd under the $\mathbb{Z}_2$ transformation, {\it CP-odd}. The $\mathbb{Z}_2$ symmetry of the singlet field is required for the stability of the DM candidate $s_I$.} $H \to H^\dagger$, $S \to S^\dagger$, and scale invariant at tree level is
\begin{eqnarray}
    V &=& \lambda_{H} \abs{H}^{4}+ \lambda_{S} \abs{S}^{4} + \frac{ \lambda'_{S} }{2} \left[ S^{4} + (\hc{S})^{4} \right] \nonumber\\
    &&+ \frac{ \lambda''_{S} }{2} \abs{S}^{2} \left[ S^{2} + (\hc{S})^{2} \right] + \lambda_{SH} \abs{S}^{2} \abs{H}^{2} \nonumber\\
    &&+ \frac{ \lambda'_{SH} }{2} \abs{H}^{2} \left[ S^{2} + (\hc{S})^{2} \right].
\label{eq:V:SM:singlet}
\end{eqnarray}
The same model has been studied with different motivation in \cite{AlexanderNunneley:2010nw,Ishiwata:2011aa,Farzinnia:2013pga}.

Of course one can also write further terms involving the combination $S +S^\dagger$, but they can be absorbed into a redefinition of parameters.
It is convenient to rewrite the scalar potential in terms of the physical fields,
\bea
 V &=& \frac14 \lambda_H \phi^4 + \frac14 \lambda_I s_I^4 +\frac{1}{4} \lambda_{RI} s_I^2 s_R^2 \nonumber\\
      &&+ \frac14 \lambda_R s_R^4 +  \frac{1}{4} \lambda_{IH} \phi^2 s_I^2 + \frac{1}{4} \lambda_{RH} \phi^2 s_R^2,
\eea
where $\phi$ is the physical  Higgs field, $s_R$ and $s_I$ are the real and imaginary parts of the singlet $S$, and \bea
 \lambda_R &=& \lambda_S+\lambda_S'+\lambda_S'' ,\\
 \lambda_I &=& \lambda_S+\lambda_S'-\lambda_S'' ,\\
 \lambda_{RI} &=& 2 (\lambda_S-3 \lambda_S') ,\\
 \lambda_{RH} &=& \lambda_{SH}+ \lambda_{SH}' ,\\
 \lambda_{IH} &=& \lambda_{SH}- \lambda_{SH}' .
\eea

The one-loop renormalization group equations  of the scalar couplings are
\begin{align}
16 \pi^2 \beta_{\lambda_H}
& = \frac{3}{8} (3 g^4 + 2 g^2 g^{\prime 2} + g^{\prime 4}) + \frac{1}{2} (\lambda_{RH}^2 + \lambda_{IH}^2) \nonumber\\
& + 24 \lambda_{H}^{2} - 3 \lambda_H (3 g^2 +  g^{\prime 2} - 4 y_{t}^2)  - 6 y_{t}^4,\label{eq:beta_lambda_H} \\
16 \pi^2 \beta_{\lambda_{R}} &= 18 \lambda_R^2 + 2 \lambda_{RH}^2 + \frac12 \lambda_{RI}^2,\label{eq:beta_lambda_R} \\
16 \pi^2 \beta_{\lambda_{I}} &= 18 \lambda_I^2 + 2 \lambda_{IH}^2 + \frac12 \lambda_{RI}^2,\label{eq:beta_lambda_I} \\
16 \pi^2 \beta_{\lambda_{RI}} &= 4 \lambda_{IH} \lambda_{RH} + 6 \lambda_{RI} (\lambda_I
+ \lambda_R) + 4 \lambda_{RI}^2, \\
16 \pi^2 \beta_{\lambda_{RH}} &= -\frac32 \lambda_{RH} ( g^{\prime 2} + 3 g^2 - 4 y_t^2) + \lambda_{IH} \lambda_{RI} \nonumber\\
 &+ 6 \lambda_{RH} ( 2 \lambda_H + \lambda_R) + 4 \lambda_{RH}^2 , \\
16 \pi^2 \beta_{\lambda_{IH}} &= -\frac32 \lambda_{IH} ( g^{\prime 2} + 3 g^2 - 4 y_t^2) + \lambda_{RH} \lambda_{RI} \nonumber\\
&+ 6 \lambda_{IH} ( 2 \lambda_H + \lambda_I) + 4 \lambda_{IH}^2 .
\end{align}

We will now see how the vacuum expectation value for $s_R$ is generated via dimensional transmutation and how it is transmitted to the SM. As in \cite{Hambye:2013dgv}, the one-loop potential can be approximated just by using a running $\lambda_R$ in the tree-level potential. We can approximate $\lambda_R$ by
\be
 \lambda_R \simeq \beta_{\lambda_R} \ln \frac{|s_R|}{s_0},
\ee
where $\beta_{\lambda_R}$ is the (always positive) beta function of $\lambda_R$, and $s_0$ is the scale at which $\lambda_R$ becomes negative.
The real part of $s$, $s_R$ and the Higgs scalar get VEVs
\be
 v = v_R \sqrt{\frac{\abs{\lambda_{RH}}}{2 \lambda_H}}, \qquad v_R \simeq s_0 e^{-1/4},
\label{eq:VEVs}
\ee
where $v$ is the SM Higgs VEV and $\lambda_{RH}<0$.

In the basis $(\phi, s_R)$ the square mass matrix for CP-even fields is given by
\be
 \left(
\begin{array}{cc}
 2 v^2 \lambda_H & -\sqrt{2} v^2 \sqrt{\lambda_H  \abs{\lambda_{RH}}} \\
 -\sqrt{2} v^2 \sqrt{\lambda_H \abs{\lambda_{RH}}} &  \abs{\lambda_{RH}} v^2+\frac{2 \beta_{\lambda_R} \lambda_H  v^2}{\abs{\lambda_{RH}}} \\
\end{array}
\right).
\label{eq:massmatrix}
\ee
 In case of small mixing (small $\lambda_{RH}$) we obtain the following eigenvalues
\bea
 m_h^2 &\simeq& v^2 \left(2 \lambda_H-\frac{\lambda_{RH}^2}{\beta_{\lambda_R}}\right), \label{mhappv1}\\
 m_s^2 &\simeq& v^2 \left(\frac{2 \beta_{\lambda_R} \lambda_H}{\abs{\lambda_{RH}}}+\frac{\lambda_{RH}^2}{\beta_{\lambda_R}}+\abs{\lambda_{RH}} \label{msappv1}\right),
\eea
while the CP-odd scalar mass is given by
\be
 m_A^2 \simeq v^2 \left(\frac{\lambda_H \lambda_{RI}}{\abs{\lambda_{RH}}}+\frac{\lambda_{IH}}{2}\right).
\ee
Eqs. (\ref{mhappv1}) and (\ref{msappv1}) are valid only if $\frac{\lambda_{RH}^2}{\beta_{\lambda_R}} \ll 1$. If this is not true,
the proper approximation is
\bea
 m_h^2 &\simeq& v^2 \left(2 \lambda_H+\abs{\lambda_{RH}}+\beta_{\lambda_R}\right) \label{mhappv2},\\
 m_s^2 &\simeq& v^2 \left(\frac{2 \beta_{\lambda_R} \lambda_H}{\abs{\lambda_{RH}}}+\beta_{\lambda_R} \label{msappv2}\right),
\eea
which usually means that the real singlet $s_R$ is lighter than the Higgs boson. The CP-even singlet $s_R$ decays to SM particles via the mixing with the Higgs boson, but the CP-odd component of the complex singlet turns out to be stable due to CP conservation, and will play the role of the DM candidate in the present SM extension. The branching ratios of the kinematically allowed decay channels of the singlet with a given mass are the same as for a SM Higgs with the same mass, and the production cross section is given by the SM Higgs production cross section multiplied by $\sin^2\theta_{SH}$, where $\theta_{SH}$ is the mixing angle between the singlet and the Higgs, obtained by diagonalising the mass matrix (\ref{eq:massmatrix}).  In the case of light $s_R$ the most constraining experimental limits are from LEP \cite{Barate:2003sz,Abbiendi:2002qp}. In the whole range below 114 GeV the upper limit for the production cross section is above $10^{-2}$ times the SM Higgs value, implying that a mixing below $\sin\theta_{SH} < 0.1$ is allowed everywhere. If the singlet mass is above the LEP reach, the constraints are from the LHC, implying $\sin^2\theta_{SH}\lesssim 0.1$ for $m_s \lesssim 500$ GeV \cite{Chatrchyan:2013yoa}. Indirect constraints on Higgs mixing from global fits of all LHC and Tevatron data imply less stringent
constraints~\cite{Giardino:2012dp,Espinosa:2012vu,Giardino:2013bma}.
In this work we will consider the limit of small mixing.

Let us now discuss the roles of the various couplings of the scalar sector in removing the global minimum of the SM Higgs potential and generating the EW symmetry breaking minimum. As described above, we start by looking at the running of $\lambda_R$. We set $\lambda_R$ to a small negative value at the EW scale. Since the beta-function (\ref{eq:beta_lambda_R}) is always positive, $\lambda_R$ will grow when running towards higher energy and will cross zero at some scale $s_0$ above the EW scale. This scale is set by the initial value of $\lambda_R$ at the EW scale and by the slope of the running set by the beta-function. Since $\lambda_R$ itself has to be small near the scale $s_0$, and since $\lambda_{RH}$ is required to be small in order to keep the mixing between $s_R$ and the Higgs small, the beta-function (\ref{eq:beta_lambda_R}) is dominated by $\lambda_{RI}$ at low scales. In order to avoid a huge hierarchy between $s_0$ and the EW scale, the running of $\lambda_R$ has to be sufficiently rapid, implying that $\lambda_{RI}$ can not be very small. Practically, to obtain $s_0$ in the range $s_0 \lesssim 10^5$ GeV, we need $\lambda_{RI}\gtrsim 0.3$ if $\lambda_R \sim - 10^{-3}$ at the EW scale. The required running from $\lambda_R$ can however be reduced by fine-tuning the starting value of $\lambda_R$ closer to zero.

To remove the global minimum of the SM Higgs potential we need to add a positive term to the beta-function of $\lambda_H$ to keep it from crossing zero. From (\ref{eq:beta_lambda_H}) we see that this can be achieved by the term $\lambda_{RH}^2 + \lambda_{IH}^2$. Since $\lambda_{RH}$ is small to avoid large mixing, this term is dominated by $\lambda_{IH}$. Thus, to remove the global minimum, we need to set a sizable initial value for $\lambda_{IH}$ at the EW scale. In practice, $\lambda_{IH}\gtrsim 0.4$ is required to keep $\lambda_H$ from running negative.

\begin{figure}[t]
\begin{center}
\includegraphics[width=0.45\textwidth]{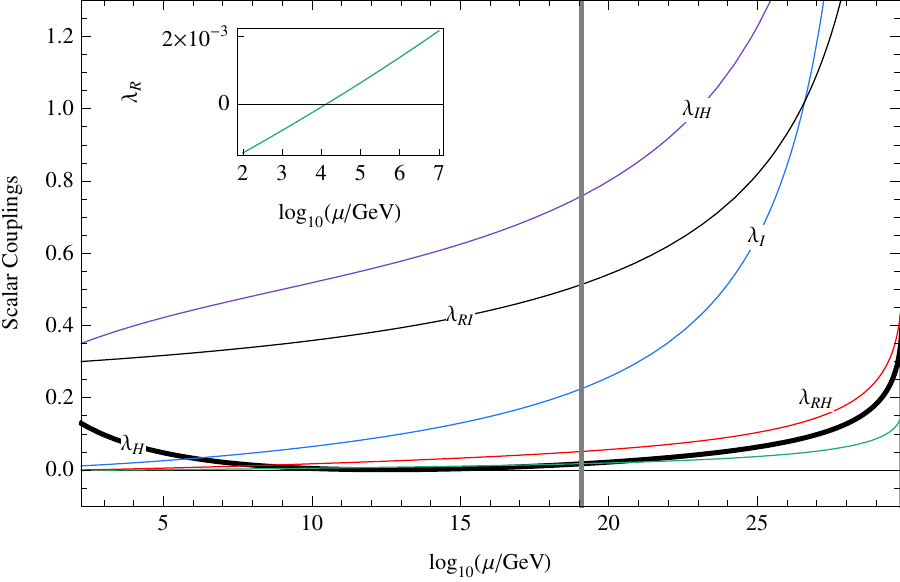}
\caption{The renormalization group running of the scalar couplings, for the initial values $\lambda_{RI} = 0.3$, $\lambda_R = -1.2 \times 10^{-3}$, $\lambda_{HI} = 0.35$, $\lambda_I=0.01$, $\lambda_{RH}=-10^{-4}$, $\lambda_H=0.12879$ and $m_t= 173.1$ GeV at the top mass scale. Notice that the Higgs self coupling remains positive in the whole range, while $\lambda_R$ runs negative around $10^4$ GeV, shown in the inset, generating a VEV at that scale.}
\label{fig:running:lambdas}
\end{center}
\end{figure}

We want to avoid generating a VEV for the imaginary part $s_I$, which means that $\lambda_I$ must stay positive. Hence we set a small positive initial value for $\lambda_I$ at the EW scale. The beta-function (\ref{eq:beta_lambda_I}) of $\lambda_I$ contains a positive contribution from both $\lambda_{IH}$ and $\lambda_{RI}$, which we know from above to have sizable values. Therefore the running of $\lambda_I$ will be quite rapid, and it will eventually run into a Landau pole, as shown in Fig.~\ref{fig:running:lambdas}. In this figure we have chosen the initial values for the parameters at the top mass scale as follows: $\lambda_{RI} = 0.3$, $\lambda_R = -1.2 \times 10^{-3}$, $\lambda_{HI} = 0.35$, $\lambda_I=0.01$, $\lambda_{RH}=-10^{-4}$, $\lambda_H=0.12879$ and $m_t=173.1$ GeV, and used beta functions at first order in the scalar couplings and second order in gauge couplings. As can be seen in the figure, the Higgs self coupling $\lambda_H$ remains positive and therefore the SM global minimum at $10^{26}$ GeV is removed, while $\lambda_R$ becomes negative around $s_0\approx 10^4$ GeV.

The position of the Landau pole of the scalar couplings depends on the choice of the initial values of the couplings at the EW scale, but it will always be below the $U(1)_Y$ Landau pole of the SM. Thus the perturbative range of our model is somewhat smaller than that of the SM without singlet. This range can, however, be easily made to extend well above the Planck scale, so for all practical purposes it makes little difference if the perturbative range of validity extends all the way up to the $U(1)_Y$ pole. Nevertheless, the validity range of the SM
is one of the results of our paper. It is needless to stress again that we do not know what happens to the SM above the Landau pole.

\begin{figure}[t]
\begin{center}
\includegraphics[width=0.45\textwidth]{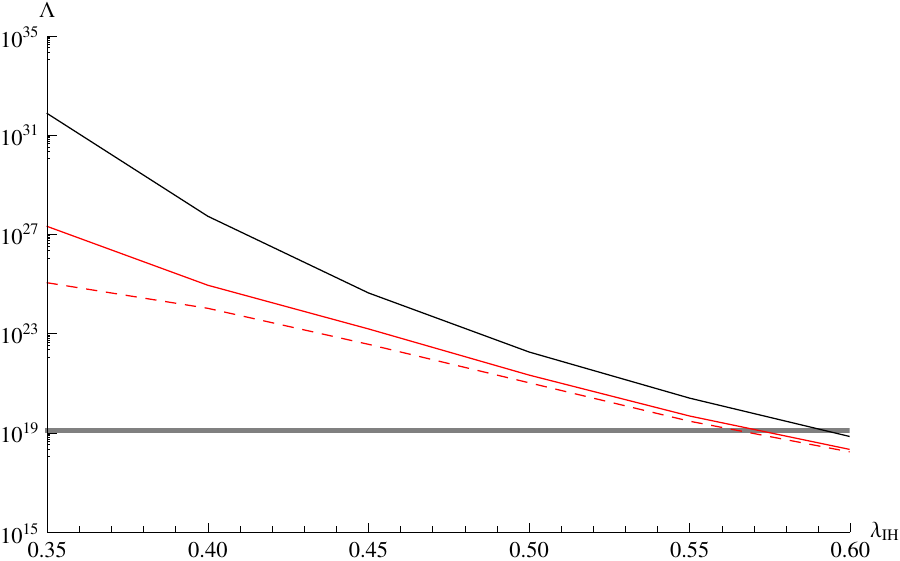}
\caption{The perturbative range of validity of the model, {\it i.e.} the position of the Landau pole of the scalar couplings, as a function of the initial value of $\lambda_{IH}$ at the EW scale. $\lambda_{RI}\simeq 0$ and  $\lambda_{R}\simeq 0$ (solid black line), $\lambda_{RI}=0.5$ and $\lambda_{R}\simeq 0$ (solid red line) and $\lambda_{RI}=0.5$ and $\lambda_{R}=0.1$ (dashed red line). The gray horizontal line is the Planck scale.}
\label{fig:scalebound}
\end{center}
\end{figure}

In Fig.~\ref{fig:scalebound} we plot the scale $\Lambda$ up to which the theory is perturbatively valid (Landau poles of the scalar quartic couplings), as a function of EW scale value of the coupling $\lambda_{IH}$ so that $\lambda_H>0$ at any scale, for given values of $\lambda_{RI}$ or $\lambda_{R}$: $\lambda_{RI}\simeq 0$ and  $\lambda_{R}\simeq 0$ (solid black line), $\lambda_{RI}=0.5$ and $\lambda_{R}\simeq 0$ (solid red line) and $\lambda_{RI}=0.5$ and $\lambda_{R}=0.1$ (dashed red line). The gray horizontal line is the Planck scale. The value of scalar Landau pole $\Lambda$ can be high,  for $0.35 \lesssim \lambda_{IH} \lesssim 0.55-0.6$ it exceeds the Planck scale, but is is always below the SM  $U(1)_Y$ Landau pole.
The reason for this behaviour is our requirement of EWSB via dimensional transmutation that severely constrains phenomenologically allowed parameter space, as described above.

It is possible to push the Landau poles of the model above the $U(1)_Y$ Landau pole of the SM, by abandoning the requirement of classical scale invariance. Then we can include tree level mass terms for the Higgs potential and for the singlet fields, and thus there is no need to generate the electroweak scale from dimensional transmutation. Therefore we no longer need to organise the running of the scalar couplings so that $\lambda_R$ crosses zero at $s_0$, and thus the requirements for the the values of the different couplings, as presented above, no longer hold. We have then complete freedom to choose the couplings in such a way that the Landau poles are above the SM UV-pole, but this comes with the price of having to put in tree level mass terms by hand. Thus there is no dynamical explanation for the value of the EW scale or the DM mass. Another possibility is that the dynamics of the singlet sector are more complicated, {\it e.g.} there is a new gauge interaction that generates the VEV of the singlet dynamically. In this case there is again more freedom to choose the initial values of the scalar couplings, with the price of a less minimal model. Finally, one can alter the input values of the SM parameters used in the analysis, $m_t$ in particular, within the experimental uncertainty. If one chooses a smaller value for $m_t$, the Higgs potential becomes more stable and one needs a smaller stabilizing contribution from the singlet sector, and for a larger value of $m_t$ one needs to generate a larger effect. Varying the input values will have some effect on the numerical results of our analysis but will not significantly affect the results. We will not consider these possibilities further in this work.

\begin{figure}[t]
\begin{center}
\includegraphics[width=0.45\textwidth]{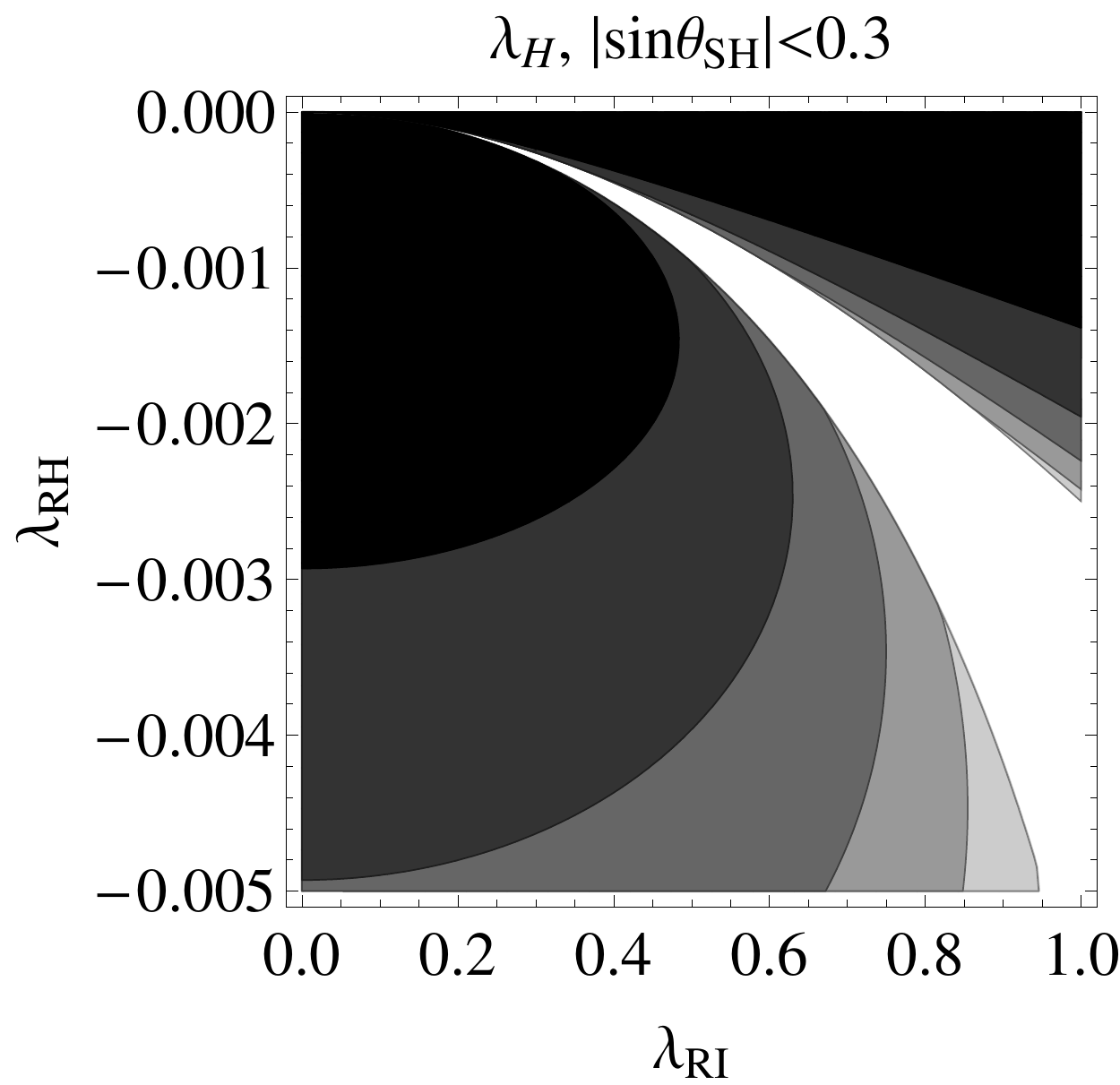}
\caption{Isocurves of $\lambda_H$ as functions of $\lambda_{RI}$ and $\lambda_{RH}$. The color scale represents the deviation $\delta\lambda_H$ from the SM value $\lambda_H\approx 0.13$. The black region corresponds to $|\delta\lambda_H|<0.001$, the darkest grey region to $|\delta\lambda_H|<0.002$ and so on, with the lightest grey region corresponding to $|\delta\lambda_H|<0.005$. The white region corresponds to large mixing, $\sin\theta_{SH}>0.3$, which we do not consider in this work. To the left of/below the white region the CP even scalar is lighter than the Higgs $m_s<m_h$. To the right/above the white region, $m_s>m_h$.}
\label{fig:LHplot}
\end{center}
\end{figure}

Finally, the initial value of $\lambda_H$ at the EW scale can be regarded as a free parameter in our model. Even though it is a SM parameter, the self coupling of the Higgs boson has not yet been directly measured at the LHC. However, the mass and the VEV of the Higgs field are known with a good precision, and in the SM they are related by
\begin{equation}
m_{H_{\rm SM}}^2=2\lambda_Hv^2,
\end{equation}
implying that the $\lambda_H$ coupling can be indirectly measured in the SM by means of this relation.
In particular, this equation fixes the value of $\lambda_H$ at the EW scale. In our model the Higgs mass is given by equation (\ref{mhappv1}) or (\ref{mhappv2}), depending on the expansion parameter as explained above. Thus the value of $\lambda_H$ deviates from the SM value by
\begin{equation}
\delta\lambda_H=\lambda_H-\lambda_{H_{\rm SM}}\simeq\frac{\lambda_{RH}^2}{2\beta_{\lambda_R}},
\end{equation}
for $\frac{\lambda_{RH}^2}{\beta_{\lambda_R}} \ll 1$. Otherwise, the deviation is given by
\begin{equation}
\delta\lambda_H\simeq -\frac{1}{2}(|\lambda_{RH}| +\beta_{\lambda_R}).
\end{equation}
In general, a large value for this deviation implies large mixing between the Higgs and the CP-even singlet $s_R$. In Fig.~\ref{fig:LHplot} we plot $\delta\lambda_H$ as function of $\lambda_{RI}$ and $\lambda_{RH}$, restricting to mixing below $\sin\theta_{SH}<0.3$. As can be seen from the figure, the deviation is typically very small. Thus, for simplicity, in the rest of the paper we will keep the initial value of $\lambda_H$ at the EW scale fixed to its SM value. This assumption has no significant effects on the results.

\section{Dark Matter}

\subsection{DM Annihilation Cross Section and Relic Density}
\begin{figure}[t]
\begin{center}
\includegraphics[width=0.45\textwidth]{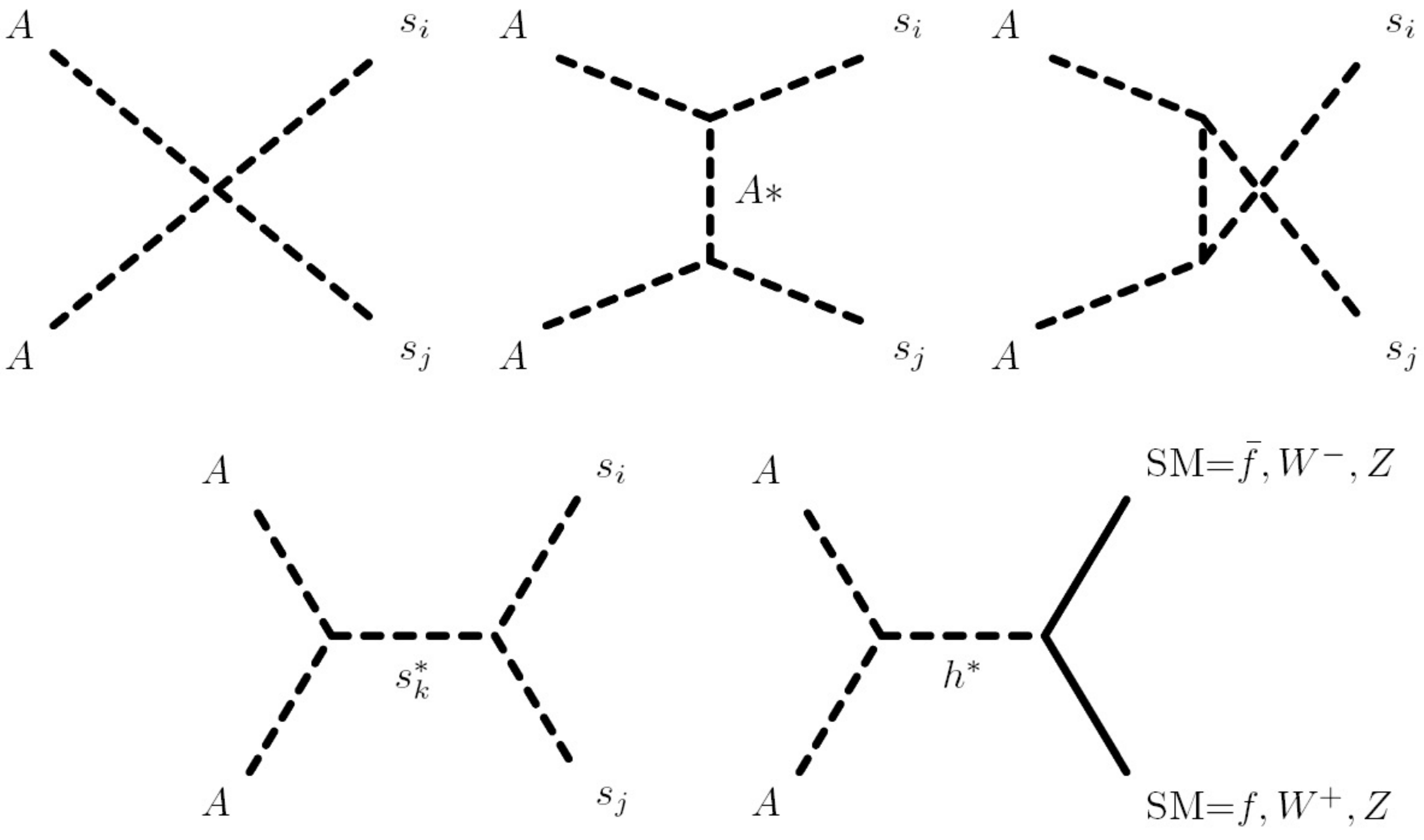}
\caption{Diagrams contributing to the DM freeze-out.} \label{fig3}
\end{center}
\end{figure}

We have extended the SM particle content with one complex singlet field $S$ without imposing any additional discrete symmetry by hand.
While the real component of $S$ acquires a VEV and triggers EWSB, the imaginary component remains stable because of the CP-invariance of the scalar potential. Therefore the corresponding scalar field will be the DM candidate of our scenario. In the following, we will use the standard notation for a pseudoscalar and denote this field by $A$.
This is the most minimal model providing dynamical EWSB and DM at the same time. Usually the stability of scalar DM is achieved with an additional $\mathbb{Z}_2$ symmetry.
An important message from our work is that this symmetry might be interpreted as CP symmetry.

In Fig.~\ref{fig3} we present the corresponding Feynman diagrams for the processes contributing to the DM freeze-out.
The DM particle $A$ can annihilate into a couple of CP-even scalars (first four diagrams in Fig.~\ref{fig3})
or into the SM particles (last diagram in Fig.~\ref{fig3}). It is known that the smallness of the doublet-singlet mixing constrains significantly the
latter processes.
In fact, the Higgs portal type DM models are already ruled out unless the main annihilation modes occur entirely in the dark
sector~\cite{Djouadi:2011aa,Greljo:2013wja}. This is the case of Dark SUSY~\cite{Heikinheimo:2013xua} and Dark Technicolour~\cite{Heikinheimo:2013fta} models.
However the constraints given in~\cite{Djouadi:2011aa,Greljo:2013wja} are valid for light DM ($m_A < 200$ GeV).
Therefore they do not apply in our present model since, as we will show in the following, we predict a relatively heavy DM, $m_A>500$ GeV.
The full cross sections corresponding to those processes are presented in the Appendix.
The relevant leading terms for the DM annihilation cross section times relative velocity are obtained by using the expansion $s \simeq 4 m_A^2 + m_A^2 v_\text{rel}^2$,
\be
\sigma_{ij} v_\text{rel} \simeq a + b \, v_\text{rel}^2,
\ee
where
\bea
a&=&
 f_{ij} \left[\sum_k \frac{a_{ijk} a_{kAA}}{4    m_A^2-m_{s_k}^2}-\frac{8 a_{iAA} a_{jAA}}{4 m_A^2-m_{s_i}^2-m_{s_j}^2}-\lambda_{ij}\right]^2 ,
\nonumber
\\
b&=& \frac{a}{16} \left(\frac{16
m_A^4-\left(m_{s_i}^2-m_{s_j}^2\right)^2}{512 \pi^2 \delta_{ij}^2 f_{ij}^2 m_A^8}-4 \right) \nonumber\\
&&- f_{ij}  \mathcal{M}_0 \left[\frac{\sum_k  a_{ijk} a_{kAA}
m_A^2}{\left(m_{s_k}^2-4 m_A^2\right)^2}   +  2 a_{iAA} a_{jAA}
F_{ij} \right], \nonumber \eea
with
\bea
 f_{ij} &=& \frac{\sqrt{1+\frac{\left(m_{s_i}^2-m_{s_j}^2\right)^2}{16 m_A^4}-\frac{m_{s_i}^2+m_{s_j}^2}{2 m_A^2}}}{8 \pi (\delta_{ij}+1) m_A^2}\nonumber ,\\
 \mathcal{M}_0 &=& 2 \left(\frac{a_{ijk}a_{kAA}}{4  m_A^2-m_{s_k}^2}+\frac{8 a_{iAA} a_{jAA}}{-4  m_A^2+m_{s_i}^2+m_{s_j}^2}-\lambda_{ij}\right) ,\nonumber\\
 F_{ij}&=&  \frac{\left(32 m_A^4-4 m_A^2 \left(m_{s_i}^2+m_{s_j}^2\right)-\left(m_{s_i}^2-m_{s_j}^2\right)^2\right)}{3 \left(-4 m_A^2+m_{s_i}^2+m_{s_j}^2\right)^3},
 \nonumber
\eea
where $m_A$ is the DM mass, $m_{s_i}$ are the CP-even scalar masses, $a_{ijk}$ the trilinear coupling of $s_i s_j s_k$
which includes also the corresponding combinatorial factor, $a_{i A A}$ the coupling of the $s_i A^2$ interaction,
$\lambda_{ij}$ the coupling of the $s_i s_j A^2$ interaction  and $v_\text{rel}$ is the relative DM velocity.
In the same way the annihilation cross section into SM particles can derived by the equations given in \cite{Guo:2010hq}.
For more details see the Appendix.

The Planck Collaboration \cite{Ade:2013lta} measured the cold DM relic density to be $\Omega_c h^2 \pm \sigma = 0.1199 \pm 0.0027$.
Since we know from experimental data $m_h \simeq 126$ GeV and $v \simeq 246$ GeV, we remain only with three relevant free parameters:
$\lambda_{RH}$, $\lambda_{IH}$ and $\lambda_{RI}$.
In order to make a first relic density estimation we consider two reference values for $\lambda_{RI}=0.5,0.02$.

\begin{figure}[t]
\begin{center}
\includegraphics[width=0.4\textwidth]{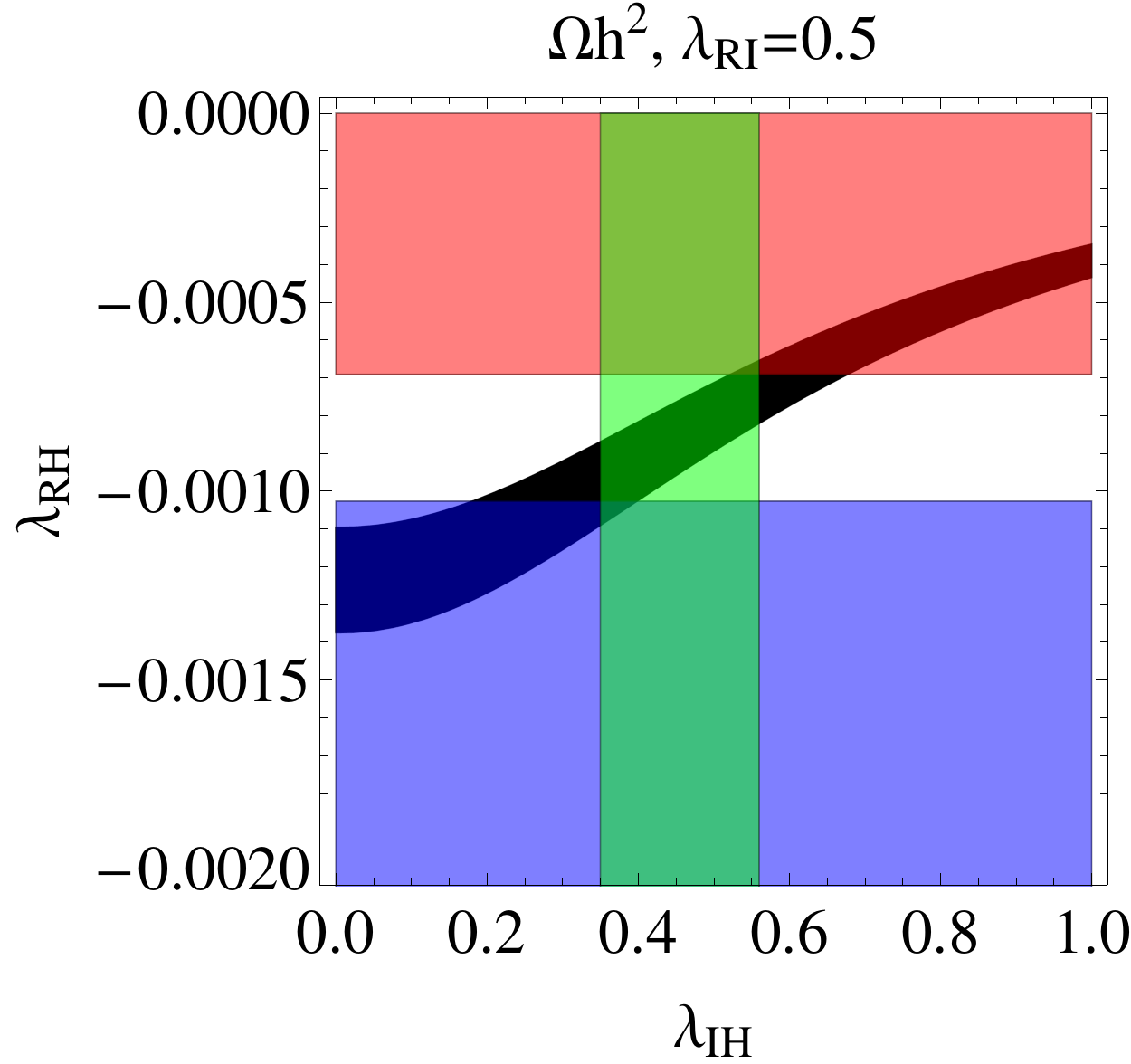}
\hspace*{1.8cm} (a)
\vskip 0.2cm
\includegraphics[width=0.42\textwidth]{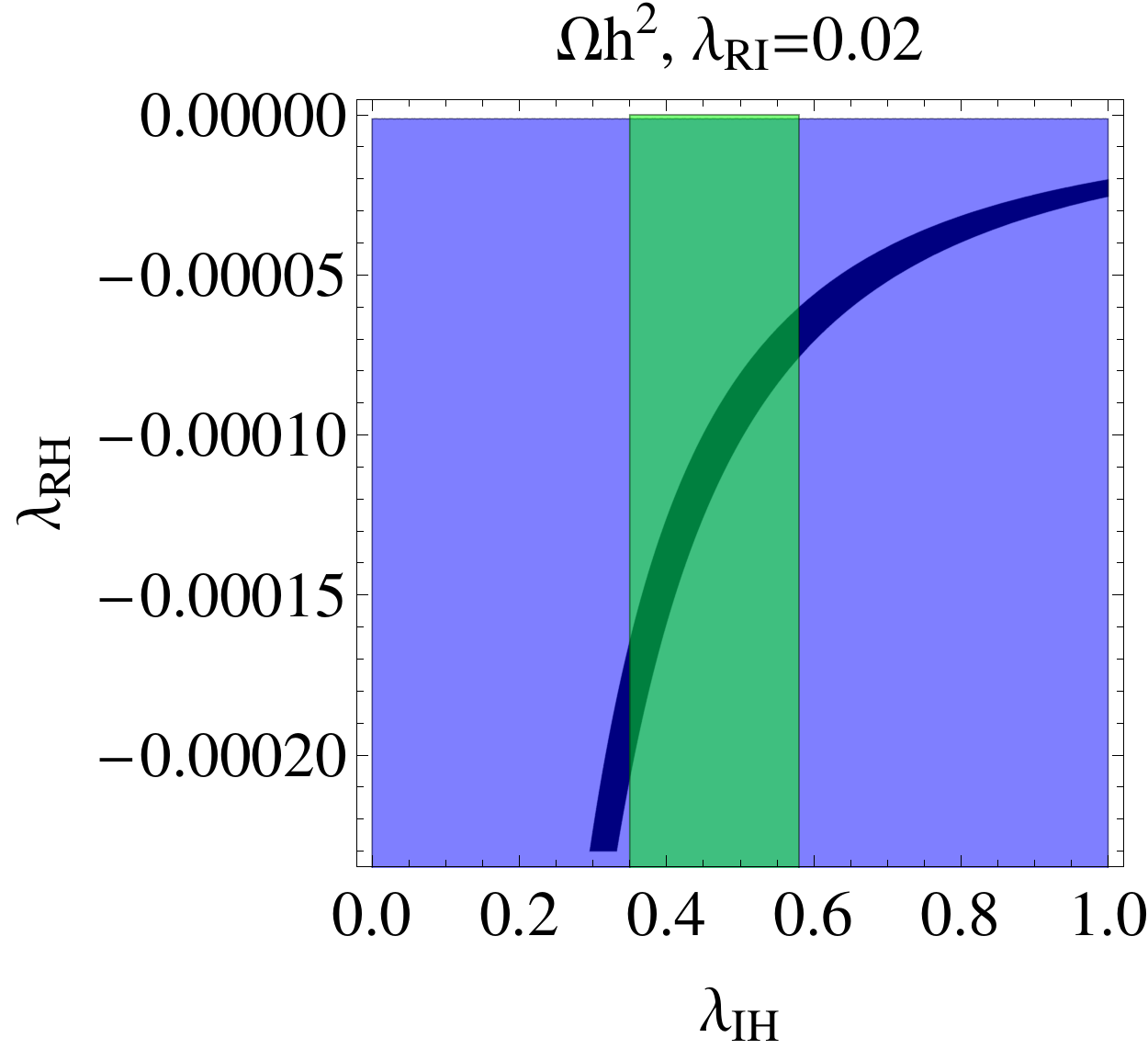}
\hspace*{1.9cm} (b)
\caption{Relic density estimates as functions of  $\lambda_{IH}$ and $\lambda_{RH}$ for  $\lambda_{RI}=0.5$ (a) and $\lambda_{RI}=0.02$ (b).
The black region corresponds to a relic density in the range $\Omega_c h^2 \pm 5 \sigma$ while
the white region is for relic densities out of the previous range.
The red region predicts Higgs boson inside the experimental bounds and $m_s > m_h$ while the blue region predicts still
Higgs boson inside the experimental bounds but $m_s < m_h$.
Finally the green region means that the validity scale of the theory is higher than Planck scale.
}
\label{fig:relic1}
\end{center}
\end{figure}

\begin{figure}[t]
\begin{center}
\includegraphics[width=0.4\textwidth]{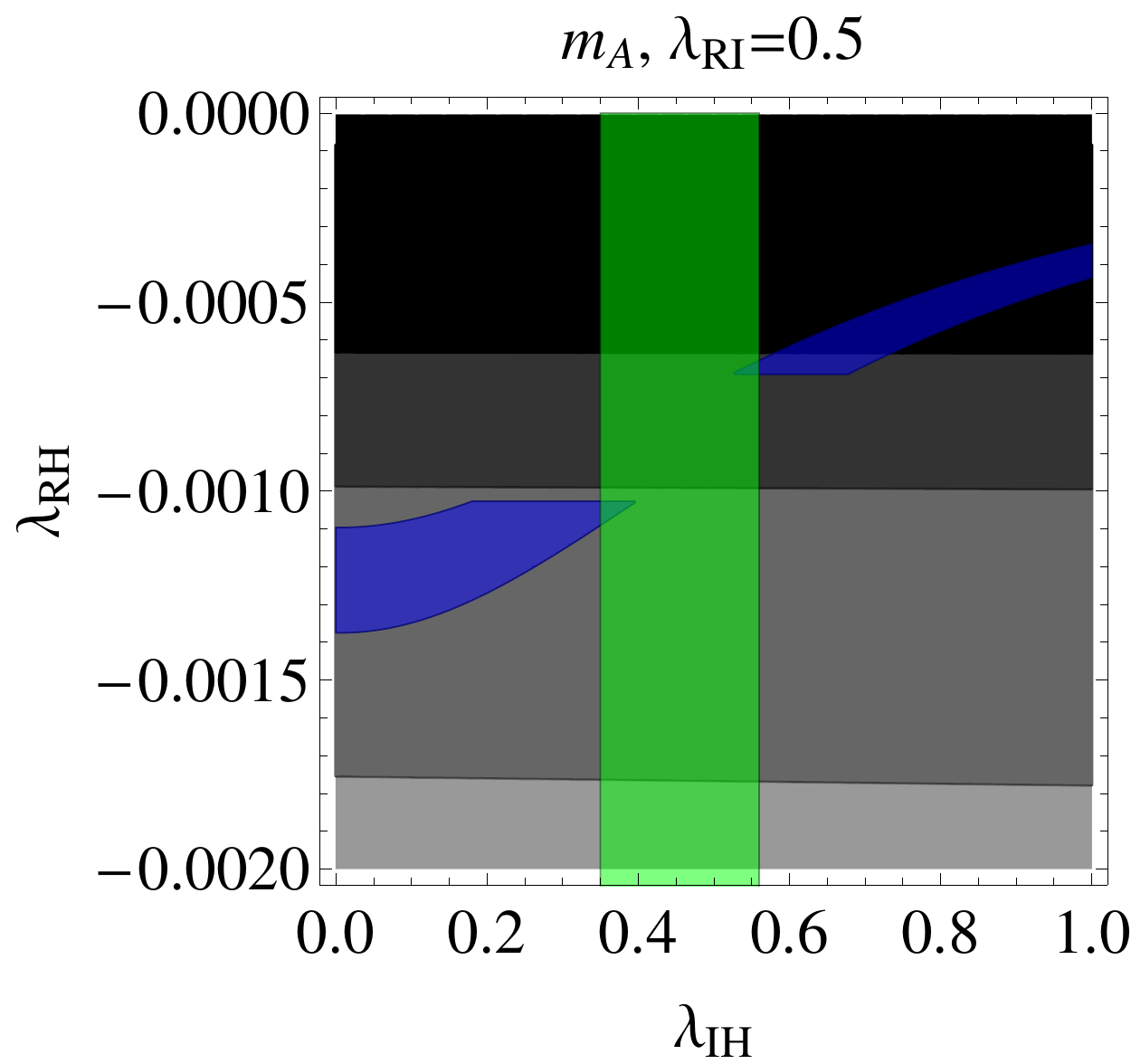}
\hspace*{1.8cm} (a)
\vskip 0.2cm
\includegraphics[width=0.42\textwidth]{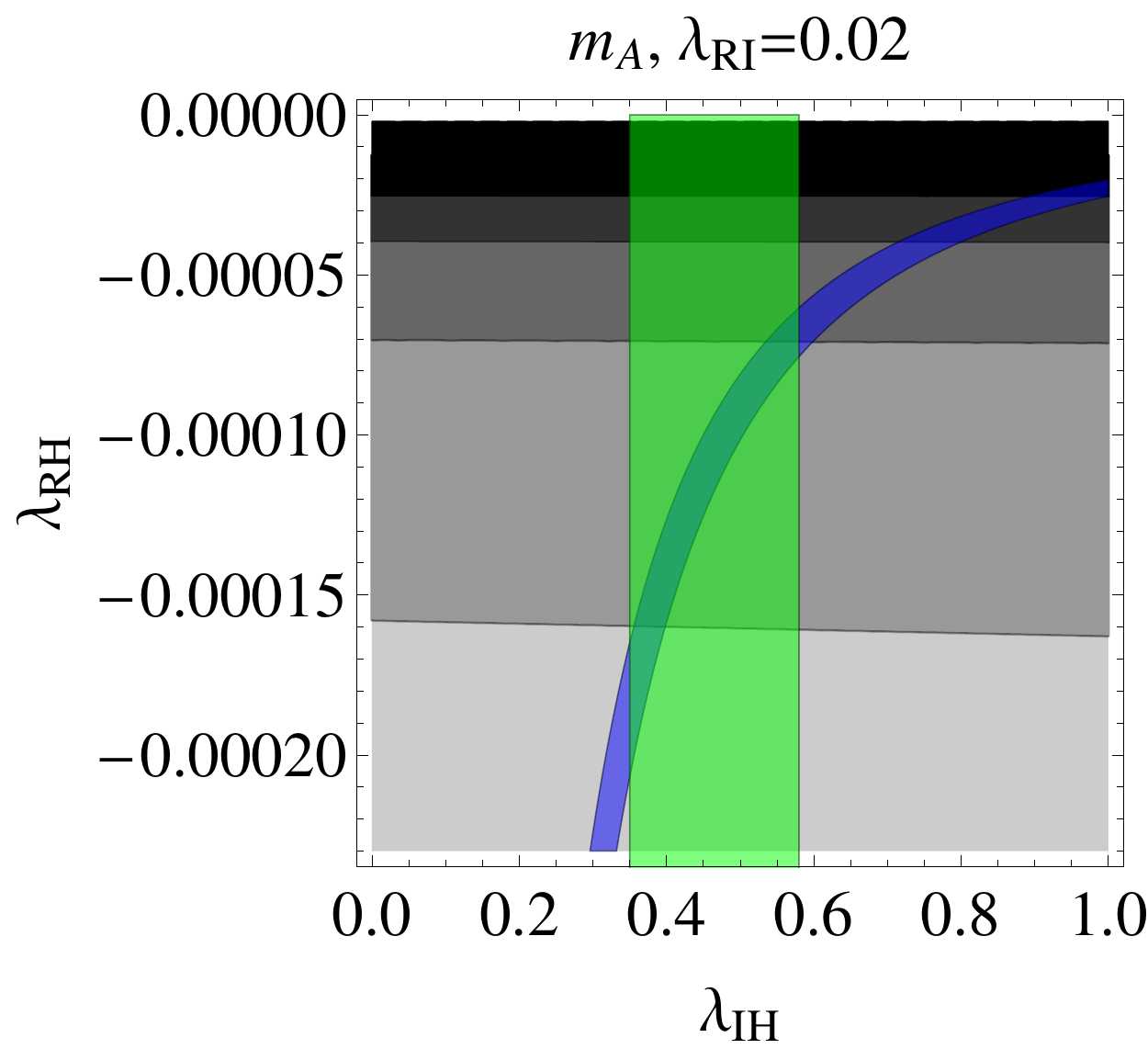}
\hspace*{1.9cm} (b)
\caption{Contour plot for $m_A$ for $\lambda_{RI}=0.5$ (a) and $\lambda_{RI}=0.02$ (b).
The black region represents masses beyond 2500 GeV then we decrease by step of 500 GeV till the lightest gray region which represents
500 GeV $<m_A<1000$ GeV.
The blue region represents the allowed region by relic density and Higgs boson mass measurements.
Finally the green region means that the validity scale of the theory is higher than Planck scale.}
\label{fig:mA1}
\end{center}
\end{figure}

Let us start with $\lambda_{RI}=0.5$.
We present our results in Fig.~\ref{fig:relic1}a in the form of a region plot as function of $\lambda_{IH}$ and $\lambda_{RH}$.
The black region corresponds to a relic density in the range $\Omega_c h^2 \pm 5 \sigma$ while the white region is for relic densities out of the previous range.
The red region predicts Higgs boson inside the experimental bounds and $m_s > m_h$ while the blue region predicts still Higgs boson inside the experimental bounds but $m_s < m_h$.
Finally the green region means that the validity scale of the theory is higher than Planck scale.
We can see that there is a wide region of parameters in agreement with present experimental data.
However if we want $\Lambda>M_\text{Planck}$, then the region is reduced to two small corners respectively
around $\lambda_{IH}\simeq 0.38$, $\lambda_{RH}\simeq -0.0011$ (with $m_s<m_h$)
and $\lambda_{IH}\simeq 0.52$ and $\lambda_{RH}\simeq -0.00065$ (with $m_s>m_h$).
In Fig.~\ref{fig:mA1}a we give a contour plot for $m_A$ for $\lambda_{RI}=0.5$ in the same ($\lambda_{IH},\lambda_{RH}$) plane.
The black region represents masses beyond 2500 GeV then we decrease by step of 500 GeV till the lightest gray region which represents
500 GeV $<m_A<1000$ GeV.
The blue region represents the allowed region by relic density and Higgs boson mass measurements.
The green region stands again for $\Lambda>M_\text{Planck}$.
We can see that in the allowed regions, $m_A \simeq$ 2-2.4 TeV.

Let us consider now $\lambda_{RI}=0.02$.
In Fig.~\ref{fig:relic1}b we present the relic density region plot as function of $\lambda_{IH}$ and $\lambda_{RH}$.
The colour code is the same as in Fig.~\ref{fig:relic1}a.
We can see that there is a wide region of parameters in agreement with present experimental data, always with $m_s<m_h$.
In Fig.~\ref{fig:mA1}b we give the contour plot for $m_A$ for $\lambda_{RI}=0.02$ in the same ($\lambda_{IH},\lambda_{RH}$) plane.
The color code is the same as in Fig.~\ref{fig:mA1}a.
We can see that in the allowed regions, $m_A \simeq$ 0.5-1 TeV.

\subsection{Direct Detection Cross Section}
\begin{figure}[t]
\begin{center}
\includegraphics[width=0.39\textwidth]{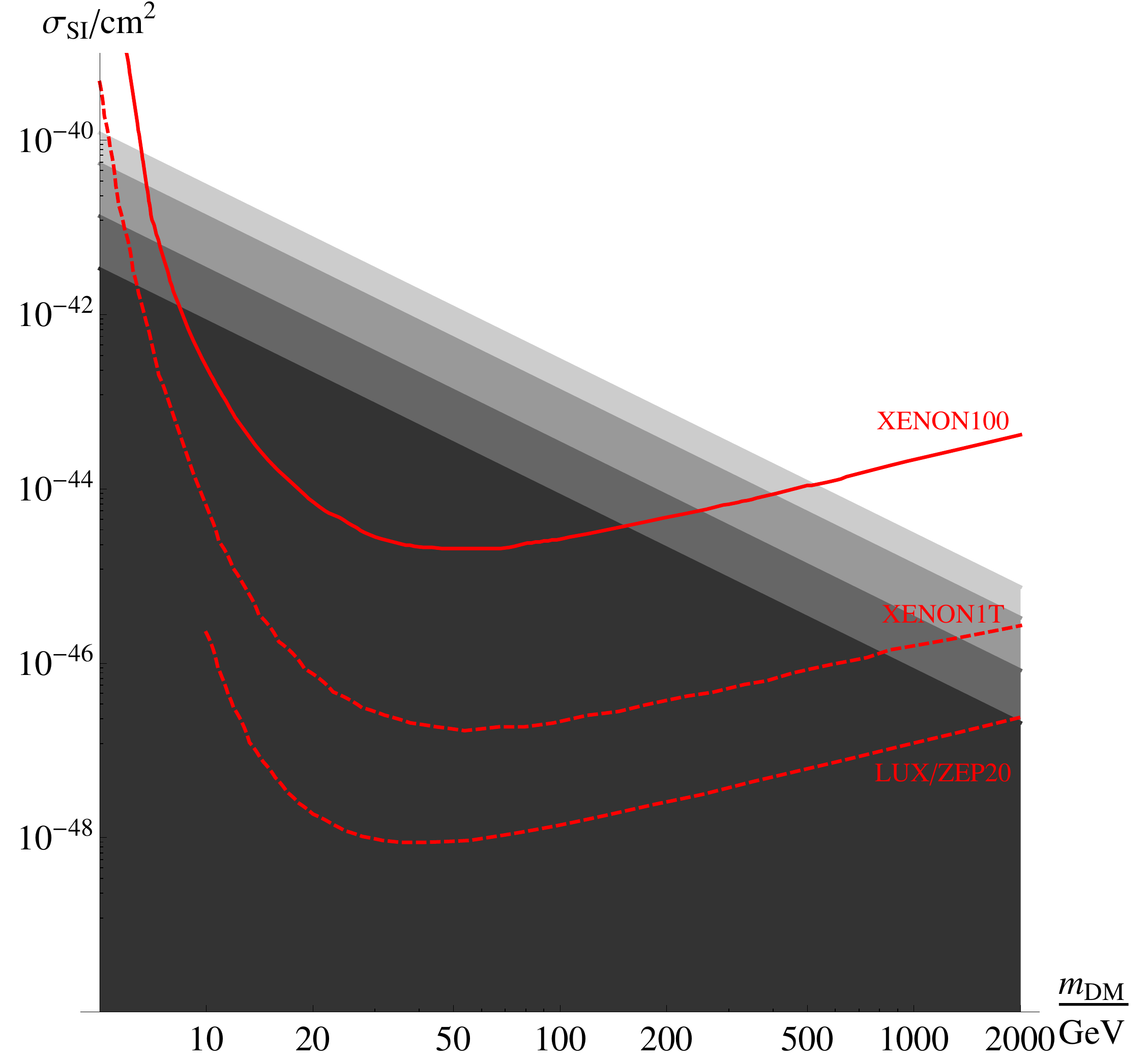}
\caption{The spin independent DM direct detection cross section for allowed parameter regions as explained in the text. The continuous red line represents present
XENON100 bound while the two dashed lines stand for XENON1T and
LUX/ZEP20 projections.} \label{fig:SIplot}
\end{center}
\end{figure}

In case of tiny mixing between the doublet and singlet scalars the spin independent DM direct detection cross section is given by
\be
 \sigma_\text{SI}^A \simeq \frac{\lambda_{AH}^2}{4\pi} \frac{m^4_N f^2}{m_A^2 m_h^4},
\ee
where $\lambda_{AH}=\lambda_{IH}$ is the quartic coupling between the CP-odd scalar and the Higgs doublet.
In Fig.~\ref{fig:SIplot} we plot our results. The shadowed gray regions represent different ranges for $\lambda_{AH}$
starting with the white region for  $\lambda_{AH}>0.6$ and continuing with
$0.4 \leq \lambda_{AH} \leq 0.6$, $0.2 \leq \lambda_{AH} \leq 0.4$ and $0.1 \leq \lambda_{AH} \leq 0.2$.
The darkest (black) region corresponds to $\lambda_{AH} \leq 0.1$.
The red continuous line represents XENON100 bound for 2012 \cite{Aprile:2012nq}, while the two red dashed lines stand for the XENON1T \cite{Aprile:2012zx}
and LUX/ZEP20  \cite{LUXZEP} projections.%
\footnote{To produce the curves, we used the online tool at http://dendera.berkeley.edu/plotter/entryform.html .}
It is evident  that the region in agreement with the relic density and Higgs boson measurements
is below the present bounds but potentially testable with the future experiments.

\section{Inflation}

As a final remark we would like to point out that the particle content of our model is also sufficient for large field chaotic inflation~ \cite{Linde:2005ht, Liddle:2000cg},
if we allow for extreme fine-tuning of the couplings\footnote{Another possibility is to consider non-minimal couplings of the scalars to gravity \cite{Bezrukov:2007ep,Bezrukov:2013fca}}. The CP-even singlet $s_R$ can act as the inflaton, if its potential is tuned to obey the slow roll conditions.
In practice this means that the quartic coupling $\lambda_R$ has to be extremely small at the Planck scale,
of the order of $\lambda_R\lesssim 10^{-13}$.
To achieve this, the initial values of the couplings at the EW scale have to be severely fine-tuned. As discussed above, the running of $\lambda_R$ is already constrained because it has to become negative at the scale $s_0$ to generate a vacuum expectation value and trigger electroweak symmetry breaking. In our discussion above, we have required $s_0$ to be reasonably close to the EW scale in order to avoid unnatural hierarchy between these scales. However, if we want to interpret $s_R$ as the inflaton field, the model will anyway contain huge fine-tuning and, therefore, we can also allow for a large hierarchy between the scale $s_0$ and the EW scale. It is then possible to tune the scalar couplings in such a way that $\lambda_R$ is negative at the EW scale, crosses zero at $s_0$ somewhere between the EW scale and the Planck scale, and remains extremely small all the way up to the Planck scale.
The chaotic inflation takes place at field values a few times  the Planck scale. According to our paradigm the SM is valid in this energy scale and the chaotic inflation
cannot be considered unnatural in our framework.

The beta-function (\ref{eq:beta_lambda_R}) of $\lambda_R$ contains positive contributions from the couplings $\lambda_{RH}$ and $\lambda_{RI}$. Therefore, to keep $\lambda_R$ extremely small all the way up to the Planck scale, also these couplings have to be very small throughout the whole range from the EW scale to the Planck scale. Fortunately there is a fixed point at $\lambda_R=\lambda_{RH}=\lambda_{RI}=0$, and thus the couplings will evolve very slowly if we tune the initial values to lie very close to this fixed point. In this limit there will be a large hierarchy between the scales $s_0$ and $\Lambda_{\rm EW}$, set by the smallness of the coupling $\lambda_{RH}$, as is apparent from equation (\ref{eq:VEVs}), and the mixing between $s_R$ and the Higgs will be extremely small, roughly
\begin{equation}
\sin\theta_{SH}\sim\sqrt{\frac{|\lambda_{RH}|}{2\lambda_H}}.
\end{equation}
Also, the inflaton $s_R$ will be very light. The lightness and the small mixing with the Higgs will reduce the decay width of $s_R$, potentially making it long lived. However, we have verified that as long as the decay channel to $e^+e^-$ is kinematically allowed, its lifetime never exceeds one second, making the scenario safe from an astrophysical point of view.
For concreteness we will give one benchmark point. We set the initial values of the couplings at the electroweak scale to $\lambda_R=-3\times10^{-14}$, $\lambda_{RI}=10^{-6}$, $\lambda_{RH}=-5\times10^{-9}$, $\lambda_{IH}=0.48$, and $\lambda_I=0.01$, while $\lambda_H$ is set to the SM value $\lambda_H\approx 0.13$. For this set of parameters the inflaton quartic coupling at the Planck scale will be $\lambda_R(M_{\rm Planck})=10^{-13}$, allowing for chaotic inflation, {\it i.e.} large field inflation \cite{Linde:1983gd}. The inflaton mass at the EW symmetry breaking vacuum is $m_s\approx 0.1$ GeV and the mixing angle is $\sin\theta_{SH}\sim10^{-4}$, yielding a lifetime of the order of $\sim 10^{-3}$ seconds for $s_R$, when the dominant decay channel is $s_R\rightarrow e^+e^-$. The production cross section for $s_R$ is $\sim 10^{-8}$ times the corresponding cross section if the SM Higgs boson mass was $0.1$~GeV, making it unobservable in collider experiments. The DM mass for the inflation benchmark point is $m_A\approx 1.3$ TeV, and the relic density is within the experimental bounds. The direct detection cross section is below the XENON100 limit, but within the projected reach of the future experiments, as shown in Fig.~\ref{fig:SIplot}.

\section{Discussion and Conclusions}

We have studied the SM in its full perturbative validity range up to  the Landau pole,
assuming that the gravity does not significantly affect the SM predictions at energies above the Planck scale. The SM without gravity can be regarded as a consistent quantum field theory all the way between \lqcd\ and the UV Landau pole of the $U(1)_Y$ gauge coupling. However, when viewed in isolation from any potential new physics, as we have assumed in this work, the SM suffers from a false vacuum problem at the EW scale, which is caused by the negative Higgs quartic coupling at an intermediate energy scale. We have proposed the most minimal extension of the SM
by one complex singlet field that solves the wrong vacuum problem, generates EWSB dynamically via dimensional transmutation,
provides the correct amount of DM, and is a candidate for the inflaton.  Compared to previous such attempts to formulate
the new SM, ours has less parameters as well as less new dynamical degrees of freedom.

In this framework the false SM vacuum is avoided due to the modification of the SM Higgs boson quartic
coupling RGE by the singlet couplings. The electroweak scale can be generated from a classically
scale invariant Lagrangian through dimensional transmutation in the scalar sector,
by letting the quartic coupling of the CP-even scalar run negative close to the
EW scale. The VEV of this scalar then induces the standard model Higgs VEV through
a portal coupling. We studied the perturbative validity range of this model and found that
the scalar quartic Landau pole appears below the SM $U(1)_Y$ Landau pole.
This happens because we demand EWSB to happen via dimensional transmutation.
If more than one singlet is added to the model, this constraint can be avoided.
Because dimensional transmutation depends only logarithmically on the energy scale,
large hierarchies can be accommodated in our model. Thus,  obtaining the right EW scale
form the high scale Landau pole is  technically natural in our framework provided that the couplings
have the right numerical values. Needless to say, we do not have any prediction why the fundamental
Yukawa and scalar self-couplings must have the needed values. In order for our model to work, some of the scalar couplings
at the EW scale have to be as small as $10^{-4}$ to provide the correct EW scale. We here simply
remind the reader that couplings of this order are already present in the SM in the form of Yukawa couplings.
Anthropic selection might be a possibility to explain the smallness of those couplings, if a suitable measure on the space of couplings can be defined.
For a recent discussion of fine-tuning in a similar model framework we refer the reader to \cite{Buttazzo:2013uya}.

The model also naturally provides a DM candidate in the form of the CP-odd scalar that is stable due to the CP-invariance of the scalar potential.
 We demonstrated that this model allows the DM particle to
be produced with the correct relic density while fulfilling all experimental constraints on Higgs boson and DM phenomenology.
Detecting the DM directly at colliders is very challenging due to the small mixing between the Higgs doublet and the singlet.
However, this framework is potentially testable in the planned DM direct detection experiments.

 We also  demonstrated that inflation can be accommodated in this model without introducing additional degrees of freedom.
 In this case the scalar couplings must be very finely tuned. Our framework does not differ from generic large scale inflation models in that
 respect.

 Our SM model extension does not provide a complete solution to the known open questions in particle physics.  Obviously, there is no model of gravity in our framework that could support our initial assumptions and explain the observed cosmological constant value. We simply assume that the presently unknown UV theory of gravity does not spoil our assumptions.
 Recent theoretical developments may support this view on gravity.
The baryon asymmetry of the universe also
requires additional dynamics, which we do not discuss. Leptogenesis remains the favourite  candidate mechanism and can easily be incorporated in our framework together with
neutrino masses.
In the context of particle physics, the strong CP problem remains unexplained, and
likely requires additional degrees of freedom to be added to this minimal model.
Clearly our results and conclusions remain valid under the assumption that these new degrees of freedom somehow decouple from the relevant degrees of freedoms that contribute to our scalar sector.

Finally we want to remark that even if the Planck scale is indeed a physical cutoff for the
validity of the Standard Model, our conclusions remain mostly valid. The extra scalars
would still avert the metastability problem of the EW vacuum, and the low
energy phenomenology of the model, including the dynamical generation of the EW scale and the
DM model, remains intact. If our framework turns out to be the right approach for
extending the validity of the SM above the Planck scale, there are concrete predictions of
our model that could be tested by future DM and collider experiments.

\vskip 1cm
\mysection{Acknowledgement}
We thank A. Strumia for discussions.
This work was supported by the ESF grants 8499, 8943, MTT8, MTT59, MTT60, MJD140, MJD435, MJD298,
by the recurrent financing SF0690030s09 project and by the European Union through the European Regional Development Fund.



\newpage

\onecolumngrid
\appendix

\section{Dark Matter Annihilation Cross Sections}
Here we give more details on the dark matter annihilation cross sections corresponding to the diagrams in Fig.~\ref{fig3}. The annihilation cross section into two scalars is given by
\bea
 \sigma_{ij} &=&\frac{p_f}{4 \pi  (\delta_{ij}+1) s \sqrt{s-4 m_A^2}} \times \\
&&\left[ 2 \left(\lambda_{ij}-\sum_k \frac{a_{ijk} a_{kAA}}{s-m_{s_k}^2}\right)^2 \right.+\frac{64 (a_{iAA} a_{jAA})^2}{\left(s - m_{s_i}^2 - m_{s_j}^2\right)^2-4  p_f^2 \left(s-4 m_A^2\right)} \nonumber\\
&&+\left.\frac{16 a_{iAA} a_{jAA} \text{arctanh}\left(\frac{2 p_f\sqrt{s-4 m_A^2}}{s - m_{s_i}^2 - m_{s_j}^2}\right) }{p_f\sqrt{s-4 m_A^2}}\left(\frac{2 a_{iAA} a_{jAA}}{s - m_{s_i}^2 - m_{s_j}^2}-\sum_k \frac{a_{ijk} a_{kAA}}{s-m_{s_k}^2}+\lambda_{ij}\right)
  \nonumber \right],
\eea
where
\be
 p_f =\frac{1}{2}\sqrt{s\left(1+\frac{\left(m_{s_i}^2-m_{s_j}^2\right)^2}{s^2}-2\frac{m_{s_i}^2+m_{s_j}^2}{s}\right)},
\ee
and
$\sqrt{s}$ is the total energy in the center of mass frame.
For what concerns the SM final states, the exact cross sections are given in \cite{Guo:2010hq}.
The relevant leading terms are
\begin{align}
\sigma_{f \bar f} v_\text{rel} & \quad \simeq \quad   \frac{\lambda_{IH}^2 m_f^2 \sqrt{1-\frac{m_f^2}{m_A^2}}}{ \left(\left(m_h^2-4 m_A^2\right)^2+\Gamma_h^2 m_h^2\right)} \times \nonumber\\
& \left[ \frac{\left(1-\frac{m_f^2}{m_A^2}\right)}{4 \pi}+
v_\text{rel}^2 \left(\frac{\left(2 m_A^2+m_f^2\right)}{32 \pi  m_A^2}
    -\frac{ \left(1-\frac{m_f^2}{m_A^2}\right) \left(48 m_A^4-16 m_A^2 m_h^2+m_h^4+\Gamma_h^2 m_h^2\right)}{16 \pi  \left(\left(m_h^2-4 m_A^2\right)^2+\Gamma_h^2 m_h^2\right)}\right)\right],
\end{align}
for the $f \bar f$ final states,
\bea
 \sigma_{W^+ W^-} v_\text{rel}  &\simeq& \frac{\lambda_{IH}^2  }{\left(\left(m_h^2-4 m_A^2\right)^2+\Gamma_h^2 m_h^2\right)} \times \nonumber\\
&&\left[\frac{\sqrt{1-\frac{m_W^2}{m_A^2}}  \left(4 m_A^4-4 m_A^2 m_W^2+3 m_W^4\right)}{8 \pi  m_A^2}\right.+
v_\text{rel}^2 \left(\frac{ \left(16 m_A^6-20 m_A^4 m_W^2+4 m_A^2   m_W^4+3 m_W^6\right)}{64 \pi  m_A^3   \sqrt{m_A^2-m_W^2}}\right.\nonumber\\
&&\left. \left.-\frac{ \sqrt{1-\frac{m_W^2}{m_A^2}} \left(4 m_A^4-4   m_A^2 m_W^2+3 m_W^4\right) \left(48 m_A^4-16   m_A^2 m_h^2+m_h^4+\Gamma_h^2   m_h^2\right)}{32 \pi  m_A^2 \left(\left(m_h^2-4 m_A^2\right)^2+\Gamma_h^2 m_h^2\right)}\right)\right],
\eea
for the $W^+ W^-$ final state, and
\bea
 \sigma_{ZZ} v_\text{rel} &\simeq&\frac{\lambda_{IH}^2  }{2\left(\left(m_h^2-4 m_A^2\right)^2+\Gamma_h^2 m_h^2\right)} \times \nonumber\\
&&\left[\frac{\sqrt{1-\frac{m_Z^2}{m_A^2}}  \left(4 m_A^4-4 m_A^2 m_Z^2+3 m_Z^4\right)}{8 \pi  m_A^2}\right.
+v_\text{rel}^2 \left(\frac{ \left(16 m_A^6-20 m_A^4 m_Z^2+4 m_A^2   m_Z^4+3 m_Z^6\right)}{64 \pi  m_A^3   \sqrt{m_A^2-m_Z^2}}\right.\nonumber\\
&&\left. \left.-\frac{ \sqrt{1-\frac{m_Z^2}{m_A^2}} \left(4 m_A^4-4   m_A^2 m_Z^2+3 m_Z^4\right) \left(48 m_A^4-16   m_A^2 m_h^2+m_h^4+\Gamma_h^2   m_h^2\right)}{32 \pi  m_A^2 \left(\left(m_h^2-4 m_A^2\right)^2+\Gamma_h^2 m_h^2\right)}\right)\right],
\eea
for the $ZZ$ final state.

\end{document}